\documentclass[12pt]{article}

\newcommand{\be}{\begin{equation}}
\newcommand{\ee}{\end{equation}}
\newcommand{\bea}{\begin{eqnarray}}
\newcommand{\eea}{\end{eqnarray}}
\newcommand{\hX}{\hat{X}}
\newcommand{\tr}{\mbox{Tr}}
\newcommand{\str}{\mbox{STr}}
\newcommand{\tg}{\tilde{\gamma}}
\newcommand{\rsl}{r'\!\!\!\!/}
\newcommand{\sym}{\mbox{Sym}}
\newcommand{\Thm}{\Theta_-}
\newcommand{\Thmd}{\Theta_-^\dagger}
\newcommand{\thm}{\theta_-}
\newcommand{\thp}{\theta_+}
\newcommand{\thmd}{\theta_-^\dagger}
\newcommand{\thpd}{\theta_+^\dagger}
\newcommand{\go}{\gamma^0}
\newcommand{\ga}{\gamma^a}
\def\abstract#1{\begin{center}{\large ABSTRACT}\end{center} \par #1}
\def\title#1{\begin{center}{\Large\bf {#1}}\end{center}}
\def\author#1{\begin{center}{\large #1}\end{center}}
\def\address#1{\begin{center}{\it #1}\end{center}}

\newcommand{\EQ}{\begin{equation}}
\newcommand{\EN}{\end{equation}}
\newcommand{\EQA}{\begin{eqnarray}}
\newcommand{\EQN}{\end{eqnarray}}
\newcommand{\EQAN}{\begin{eqnarray*}}
\newcommand{\EQNN}{\end{eqnarray*}}

\renewcommand{\theequation}{\arabic{section}.\arabic{equation}}
\newcommand{\Tr}{{\rm Tr}}
\def\identity{{\rlap{1} \hskip 1.6pt \hbox{1}}}

\makeatletter
\@addtoreset{equation}{section}
\makeatother
 
\begin{document}
\begin{titlepage}
\hspace*{\fill}
\vbox{
\hbox{hep-th/0207197}
\hbox{KEK-TH-836}
\hbox{TIT/HEP-481}
\hbox{July, 2002}}
\vspace*{\fill}
\begin{center}
 \Large\bf 
Supersymmetric action of multiple D0-branes\\
 from Matrix theory 
\end{center}
\vskip 1cm
\author{
Masako Asano \footnote{E-mail address:\ \ 
{\tt asano@post.kek.jp}}}
\address{
Theory Division, 
Institute of Particle and Nuclear Studies\\
KEK, High Energy Accelerator Research Organization\\
Tsukuba, Ibaraki, 305-0801 Japan}

\begin{center} and \end{center}

\author{
Yasuhiro Sekino \footnote{ 
E-mail address:\ \ {\tt sekino@th.phys.titech.ac.jp}}}
\address{
Department of Physics,  Tokyo Institute of Technology, \\
Ookayama, Meguro-ku, Tokyo 152-8550 Japan}

\vspace{\fill}
\abstract{We study one-loop effective action of
Berkooz-Douglas Matrix theory and 
obtain non-abelian action of D0-branes in the
longitudinal 5-brane background. In this paper, we
extend the analysis of hep-th/0201248 and 
calculate the part of the effective action containing 
fermions. We show that the effective action is manifestly
invariant under the loop-corrected SUSY transformation,
and give the explicit transformation laws.
The effective action consists of blocks which are
closed under the SUSY, and it includes  the supersymmetric
completion of the couplings to the longitudinal 5-branes
proposed by Taylor and Van Raamsdonk as a subset.}

\vspace*{\fill}


\end{titlepage}

\section{Introduction}
 
The fact that multiple D-branes are described by the
matrix valued coordinates led to many surprising effects.
Especially, D$p$-branes are allowed to
couple to Ramond-Ramond forms of higher degree than $p+1$ 
\cite{TR1,My}. 
Due to those couplings, non-commutative configurations are
stabilized in the presence of certain background fields,
which are interpreted as the D-branes having finite
extent \cite{My}. To study such intrinsically stringy effects, 
precise understanding of the effective action of multiple
D-branes in background fields is needed. 

It is well known that a single D-brane in the low acceleration limit
is described by the Born-Infeld action which includes all the
$\alpha'$ corrections.
However, it is not clear how to generalize it
to the case of multiple D-branes which have non-abelian 
gauge symmetries. In the flat background,
the effective action of $N$ coincident D-branes 
is given by the maximally supersymmetric U($N$) 
Yang-Mills theory in the small $\alpha'$ limit, 
but the terms of higher orders in $\alpha'$ are not
well-understood. 

Supersymmetrization of multiple D-brane action including
these higher order terms is also a hard problem.
In the abelian case, there exists $\kappa$-symmetric 
formulation \cite{C1,Sc}
which gives a supersymmetric action after gauge 
fixing the world-volume diffeomorphism and the 
$\kappa$-symmetry \cite{Sc},
but an attempt to define $\kappa$-symmetry with non-abelian
parameter~\cite{B1} does not seem successful \cite{B2}.
Studies for obtaining  
$\alpha'$ corrections to the SYM at the first few
orders including the fermionic part is being done from various 
standpoints (see the references cited below for 
current status): i) from the calculation of open string
disk amplitudes \cite{B2}; 
ii) by constructing 
the invariants with respect to the $\alpha'$-corrected
SUSY transformation \cite{C2,dR1};
iii) by requiring the existence of certain BPS states
which are present in string theory \cite{dR2}.

When we consider multiple D-branes in non-trivial backgrounds,
determining the action is further difficult. Background fields
should be regarded as a function of matrix fields somehow,
but the principles for doing so is not clear. As for the
supersymmetrization, there is practically no knowledge 
up to now.

Effective action of 
Matrix theory \cite{BFSS} gives insight for the
multiple D-brane action. Taylor and Van Raamsdonk 
studied the effective action of BFSS Matrix theory in detail 
and found the terms which can be interpreted as
the supergravity interactions. From those terms, they
read off the coupling of D-branes to weak background
fields, and further proposed a form of the couplings
to general weak background fields \cite{KT,TR1,TR2,TR3}. 
Those couplings were applied to various contexts, 
including the gauge-theory calculation of
the absorption cross section of dilaton higher partial 
waves by D3-branes, which exactly reproduces the
semiclassical supergravity results \cite{KTR}.

The subject of our paper is  Matrix theory proposed
by Berkooz and 
Douglas \cite{BD}, which is the matrix model for M-theory
in the presence of longitudinal(L) 5-branes.
Berkooz-Douglas(BD) Matrix theory is defined by the 0-4 string
and 0-0 string sectors of the SYM
describing D0-D4 system. 
In a previous paper \cite{AS}, we performed  one-loop
integration of 0-4 fields and
obtained the bosonic part of the
non-abelian action of D0-branes. 
We found that the action consists of the terms given from
the general proposal of Taylor and Van Raamsdonk by substituting
the L5-brane backgrounds, plus the corrections involving extra 
commutators. 
Since L5-branes are degrees of freedom which are not present in BFSS
Matrix theory, the fact that the proposed couplings were exactly
reproduced is regarded as a non-trivial evidence for the
consistency of the BFSS and BD matrix models.

In this paper, extending the analysis of ref.\cite{AS},
we calculate the one-loop effective action of BD Matrix theory 
and obtain non-abelian effective action of D0-branes 
including the part containing fermionic fields. 
Especially, we reveal the consequence of the supersymmetry 
of the classical action. 
We point out that the one-loop effective action is manifestly
invariant under effective SUSY transformation.
The transformation law is given simply by the one-loop
expectation value of the classical SUSY transformation law.
By examining the transformation rules, we decompose the 
terms into the blocks which close within themselves 
under the transformation. 
Among these blocks, we identify the  supersymmetric
completion of the bosonic terms given by the Taylor 
and Van Raamsdonk's proposal
applied to the longitudinal
5-brane background.

This paper is organized as follows. In section 2, we
review the action and SUSY transformation of BD Matrix
theory. In section 3, after explaining the method of
the loop calculation, we present the results for the
fermionic terms of the effective action. In section 4,
we discuss the SUSY of the effective action.
We conclude with remarks on the directions for future
studies in section 5.
We summarize the representation
of spinors and gamma matrices adopted in this
paper in Appendix A, and derive 
the invariance of one-loop effective action
under the loop-corrected SUSY transformation
in Appendix B.

%

\section{Berkooz-Douglas Matrix theory}
The action of Berkooz-Douglas Matrix theory is given by the
0-0 and 0-4 string sectors of the SYM which describe the
D0-D4 bound state. In the case of $N$ D0-branes and
$N_4$ D4-branes, the theory has U($N$) gauge symmetry 
and U($N_4$) global symmetry (which is not gauged, for
the gauge fields on D4-branes are discarded).
The action is based on the D=6 ${\cal N}$=1 SYM
and reads as follows.
\begin{equation}
S=S_0+S_5
\label{BDaction}
\end{equation}
\begin{eqnarray}
S_0&=& {1\over g_s\ell_s}\int\! dt\, 
\Tr\, \Big( {1\over 2}D_0 X_i D_0 X_i
+{1\over 4 \lambda^2} [X_i,X_j]^2 
-i \thmd D_0 \thm -i \thpd D_0 \thp \nonumber \\
&&\hspace{2cm} +{1\over \lambda}\thmd \go \ga [\thm, X_a]
+{1\over \lambda}\thpd \go \ga [\thp, X_a]
+{1\over \lambda}\thpd \go [\thm, \phi_2]
-{1\over \lambda}\thmd \go [\thp, \bar{\phi}_2]\nonumber\\
&&\hspace{2cm}+{1\over \lambda}\thp C [\thm, \phi_1]
+{1\over \lambda}\thpd C^* [\thmd, \bar{\phi}_1] \Big)\label{S0}\\
S_5&=&{1\over g_s\ell_s} \int\! dt\, \Big( (D_0 v_I)^\dagger D_0 v_I 
-{1\over \lambda^2}v_I^\dagger (X_a)^2 v_I -i\chi^\dagger D_0 \chi 
-{1\over \lambda}\chi^\dagger \gamma^0\gamma^a X_a\chi\nonumber\\
&&\hspace{1cm}
 -{1\over 2\lambda^2}\{ v_1^\dagger ([\phi_1,\bar{\phi}{}_1] 
+[\phi_2,\bar{\phi}{}_2])
v_1 - v_2^\dagger ([\phi_1,\bar{\phi}{}_1] +[\phi_2,\bar{\phi}{}_2])
v_2 
-2v_2^\dagger ([\bar{\phi}{}_1,\bar{\phi}{}_2]) v_1 \nonumber\\
&&\hspace{1.5cm} +2v_1^\dagger ([\phi_1, \phi_2]) v_2 \} 
-{\sqrt{2}\over \lambda}\{\chi^\dagger \go \thm v_1
-\chi^\dagger C^* \thm^\dagger v_2
-v_1^\dagger \thm^\dagger \go \chi
+v_2^\dagger \thm C\chi \}\nonumber\\
&&\hspace{1cm}
-{1\over 2\lambda^2}\{(v_1^\dagger v_1)(v_1^\dagger v_1)
+(v_2^\dagger v_2)(v_2^\dagger v_2)
 -2(v_1^\dagger v_2)(v_2^\dagger v_1)
+4(v_1^\dagger v_1)(v_2^\dagger v_2) \}
\Big)
\label{S5}
\end{eqnarray}
where we use the indices $a,b=5,\ldots, 9$ for 
the space transverse to the D4-branes, $m,n=1,\ldots, 4$
for the space along the D4-branes, and 
$i,j=1,\ldots, 9$ to denote both of them. 
Dimensionful parameter $\lambda$ is defined by 
$\lambda=2\pi \ell_s^2$. 

The part $S_0$ is the terms containing only the 0-0 sector
fields ($X_a$, $\phi_1$, $\phi_2$, $\thm$, $\thp$),
which is nothing but the BFSS action, i.e. the dimensional
reduction of D=10, ${\cal N}=1$ SYM.
The 0-0 sector fields are in the adjoint rep. of
U($N$) and are singlets of U($N_4$). Covariant derivatives
for those fields are defined as $D_0 X_i=\partial_0 X_i+
i[A, X_i]$.
Note that we have defined the complex combination of
$X_m$ by $(\phi_1,\phi_2)=(X_1+iX_2,X_3+iX_4)$,
and that the fermions are expressed as
6D Weyl spinors, which have 4 complex components.
As in the previous paper
\cite{AS}, we do not use the SU(2) Majorana convention,
for we prefer unconstrained fermions for the loop
calculations.
The subscripts on the spinors $\thp$, $\thm$ denote
the positive and negative 6D chiralities, respectively
($\overline{\gamma} \theta_{\pm}=\pm \theta_\pm$,
where $\overline{\gamma}=\go \gamma^1 \ldots \gamma^5$).
The matrix $C$ is the charge conjugation matrix.
We also use the `complex conjugation matrix' $B$.
See Appendix A for our conventions
for spinors and gamma matrices, 
the relation between 10D and 6D notations, and
the definitions of $C$ and $B$.
We note the reader that in this paper,
the transpose of the spinor indices is not
indicated explicitly. 
For example, $\thp C [\thm, \phi_1] $ in eq.(\ref{S0})
means $\thp^t C [\thm, \phi_1]$ where $t$ denotes the
transpose of spinor indices (but not of gauge indices).

The additional part $S_5$ contains the 0-4 sector fields
($v_I$, $\chi$). These fields are given by
the hypermultiplets of the 6D theory which consist of 
2 complex bosons $v_I$ ($I=1,2$) 
and a complex spinor $\chi$ with positive 6D chirality
($\overline{\gamma} \chi=+\chi$).
Both of them are in the bi-fundamental rep.
of U($N$)$\times$U($N_4$).
Covariant derivatives are defined as $D_0 v_I
=\partial_0 v_I +i A v_I$.
We remark here that only half of the 0-0 sector fermions
($\thm$) couple to the 0-4 sector fields. 

This model has half the amount of supersymmetry
as the BFSS model. The SUSY parameter 
$\eta$ is a complex 6D spinor with  negative chirality, 
thus the number of (real) supercharges is 8.
The SUSY transformation law is given as follows.
The 0-0 sector fields transform as
\EQA
&&\delta A=\delta^{(0)}A=
{i\over \lambda}(\eta^\dagger \thm -\thmd \eta),\quad
\delta X_a=\delta^{(0)}X_a
 =i(\eta^\dagger\go\ga \thm -\thmd \go\ga \eta),\nonumber\\
&&\delta \phi_1=\delta^{(0)}\phi_1 =-2i \eta^\dagger C^* \thpd,\quad
\delta \phi_2=\delta^{(0)}\phi_2 = -2i \eta^\dagger \go \thp,\nonumber\\
&&\delta \thp=\delta^{(0)} \thp =D_0\bar{\phi}_1 C^* \eta^\dagger +
D_0 \phi_2 \go \eta +{i\over \lambda}[X_a,\bar{\phi}_1]
B^* \ga{}^* \eta^\dagger +{i\over \lambda}[X_a,\phi_2]\ga\eta, 
\nonumber\\
&&\delta \thm= \delta^{(0)}\thm + \delta'\thm
\label{BDSUSY00}
\EQN
where
\EQ
\delta^{(0)}\thm=
D_0 X^a \go\ga\eta +{i\over 2\lambda}
[X_a,X_b]\gamma^{ab}\eta
-{i\over 2\lambda}([\phi_1,\bar{\phi}_1]+[\phi_2,\bar{\phi}_2])\eta
+{i\over \lambda}[\bar{\phi}_1,\bar{\phi}_2]
B^* \eta^\dagger,
\label{BDSUSYthm0}
\EN
\EQ
\delta'\thm=
{i\over \lambda}(-v_1 v_1^\dagger+v_2 v_2^\dagger)\eta
-{2i\over\lambda} (v_2 v_1^\dagger) B^*\eta^\dagger.  
\label{BDSUSYthmp}
\EN
We have denoted $\delta^{(0)}$ the part of the transformation laws 
which does not contain 0-4 sector fields in the RHS. It is the 
same as the transformation law for the BFSS theory, except for
the fact that the parameter is restricted to 
$\overline{\gamma}\eta=-\eta$ now. 
Only $\delta\thm$ has extra contribution $\delta'\thm$
containing 0-4 sector fields. 
The transformation law for the 0-4 sector fields is as 
follows 
\EQA
&& \delta v_1 =i\sqrt{2} \eta^\dagger \go \chi,\quad
\delta v_2= -i\sqrt{2} \eta C \chi,\nonumber\\
&& \delta \chi =-\sqrt{2} (D_0v_2\go B^* \eta^\dagger
+D_0 v_1 \go \eta)
-{\sqrt{2}i\over \lambda} X_a\ga (v_2 B^* \eta^\dagger+v_1\eta).
\label{BDSUSY04}
\EQN
 
\section{One-loop effective action}
\subsection{Method of the perturbative calculation}
We obtain the effective action as the one-loop determinant
(Berezinian) of the fluctuations. It is obtained from the
part of the action quadratic in the fluctuations
\bea
S^{\mbox{\scriptsize(quad)}}&=&  
v^\dagger  K_{\mbox{\scriptsize bos}} v+
v^\dagger K^\dagger_{\mbox{\scriptsize mix}}\chi + \chi^\dagger K_{\mbox{\scriptsize mix}} v
+\chi^\dagger K_{\mbox{\scriptsize fermi}} \chi \nonumber\\
&=& (v^\dagger+\chi^\dagger K_{\mbox{\scriptsize mix}} K^{-1}_{\mbox{\scriptsize bos}})
K_{\mbox{\scriptsize bos}}(v+ K^{-1}_{\mbox{\scriptsize bos}} 
K^\dagger_{\mbox{\scriptsize mix}} \chi)
+\chi^\dagger (K_{\mbox{\scriptsize fermi}}-K_{\mbox{\scriptsize mix}}K^{-1}_{\mbox{\scriptsize bos}}K^\dagger_{\mbox{\scriptsize mix}}) \chi
\nonumber
\eea
as
\begin{equation}
S^{(1)}= \ln {\rm Det} K_{\mbox{\scriptsize bos}} -\ln {\rm Det} 
K_{\mbox{\scriptsize fermi}}
-\ln {\rm Det} (1- K^{-1}_{\mbox{\scriptsize fermi}}K_{\mbox{\scriptsize mix}}K^{-1}_{\mbox{\scriptsize bos}}
K^\dagger_{\mbox{\scriptsize mix}}).
\label{S1}
\end{equation}
We perform the calculations in the Euclidean formulation 
($\tau=i t$, $S^{(M)} \rightarrow S^{(E)}=i S^{(M)}$) and
rotate the result to Minkowski. Note that indices and the integration
symbols are implicit in the above equations and eq.(\ref{Gmix})
below.
The first two terms of (\ref{S1}) are the ones obtained in 
the previous paper~\cite{AS}. We calculate 
\EQA
\Gamma_{\mbox{\scriptsize mix}}&\equiv& \Tr\, \ln (1- 
K^{-1}_{\mbox{\scriptsize fermi}}K_{\mbox{\scriptsize mix}}
K^{-1}_{\mbox{\scriptsize bos}}K^\dagger_{\mbox{\scriptsize mix}})
\nonumber\\
&=&\Tr\, \sum^\infty_{\ell=1}{1\over \ell}
(K_{\mbox{\scriptsize bos}}^{-1}K_{\mbox{\scriptsize mix}}^\dagger K_{\mbox{\scriptsize fermi}}^{-1}K_{\mbox{\scriptsize mix}})^\ell
\label{Gmix}
\EQN
in this paper.

The calculation is performed using a perturbative expansion.
We first separate the backgrounds for bosonic matrices as 
\EQ
(X_0,X_m, X_a)= (\hX_0,\hX_m ,r_a+\hX_a)
\label{Xdecomp}
\EN 
where $r_a$ are constants proportional to the
identity matrix. ($X_0$ is the Euclidean continuation of
$A$ which is defined by $X_0=-i\lambda A$.)
We divide the action $S^{\mbox{\scriptsize(quad)}}$ into the 
`free' part, which does not contain $\hX$
or fermion background
$\Theta$, and the interaction vertices. 
The free part is given by
\be
S^{\mbox{\scriptsize (free)}}= {1\over g_s\ell_s}\int\! d\tau\, 
\left\{
v^{\dagger}
\left(-\partial_{\tau}^2+\frac{r^2}{\lambda^2}
\right)v +
\chi^{\dagger}\left(-\partial_{\tau}+\frac{1}{\lambda}
\tilde{\gamma}^a r_a
\right)\chi
\right\}
\label{Sfree}
\ee 
where $\tg^a = \gamma^0\gamma^a$. 
The propagators are determined from (\ref{Sfree}) as
\be
\langle v_{I,A\tilde{A}}(\tau) v_{J,B\tilde{B}}^\dagger
(\tau') \rangle
=(g_s\ell_s) \delta^{IJ}\delta^{AB}\delta^{\tilde{A}\tilde{B}}
\int\frac{dk}{2\pi}\frac{e^{ik(\tau-\tau')}}{k^2+r'^2}
\equiv (g_s\ell_s) \delta^{IJ}
\delta^{AB}\delta^{\tilde{A}\tilde{B}}
\Delta(\tau-\tau').
\ee
\begin{eqnarray}
\langle \chi_{\alpha,A\tilde{A}}(\tau) 
\chi_{\beta,B\tilde{B}}^\dagger
(\tau') \rangle
&=& (g_s\ell_s) \delta^{AB}\delta^{\tilde{A}\tilde{B}}
\int\frac{dk}{2\pi}\frac{e^{ik(\tau-\tau')}}{k^2+r'^2}
\left(r'\!\!\!\!/+ik
\right)_{\alpha\beta}\nonumber\\
&=& (g_s\ell_s) \delta^{AB}\delta^{\tilde{A}\tilde{B}}
\left(\partial_\tau+r'\!\!\!\!/
\right)_{\alpha\beta}
\Delta(\tau-\tau')
\end{eqnarray}
where $A$,$B$ are the U($N$) indices, $\tilde{A}$,$\tilde{B}$ are
the U($N_4$) indices and $\alpha,\beta,\ldots (=1,\ldots,8)$ are the
6D spinor indices. We have also defined
$r'=r/\lambda$ and $ r'\!\!\!\!/=\tg^a r_a/\lambda$.

Boson-boson vertices are
\EQ
S^{\mbox{\scriptsize (int)}}_{\mbox{\scriptsize bos}}
={1\over g_s\ell_s}
\int\! d\tau\, \frac{1}{\lambda^2}v_I^\dagger V_{IJ}(\tau)v_J 
\label{Sintbos}
\EN
where 
\EQA
&&V_{11}=(2r_a \hX_a +\hX_a^2 
+\hX_0^2-i\lambda\partial_\tau \hX_0-2i\lambda \hX_0\partial_\tau)
+{1\over 2}([\phi_1,\bar\phi_1]
+[\phi_2,\bar\phi_2])\nonumber\\
&&V_{22}=(2r_a \hX_a +\hX_a^2 
+\hX_0^2-i\lambda\partial_\tau \hX_0-2i\lambda \hX_0\partial_\tau)
-{1\over 2}([\phi_1,\bar\phi_1]
+[\phi_2,\bar\phi_2])\nonumber\\
&&V_{12}=[\phi_1,\phi_2],\quad V_{21}=-[\bar\phi_1,\bar\phi_2].
\EQN
Fermion-fermion vertices are 
\EQ
S^{\mbox{\scriptsize (int)}}_{\mbox{\scriptsize fermi}}=
{1\over g_s\ell_s}\int\! d\tau\, 
\frac{1}{\lambda}\chi^\dagger (\tilde{\gamma}^a\hX_a -i \hX_0)\chi 
\label{Sintfermi}
\EN
Boson-fermion mixed vertices are
\EQ
S^{\mbox{\scriptsize (int)}}_{\mbox{\scriptsize mix}} = 
{1\over g_s\ell_s}
 \int\! d\tau\, {\sqrt{2}\over \lambda} 
\Big\{ \chi^\dagger_\alpha L_{I\alpha} v_I
+v_I^\dagger L_{I\alpha}^\dagger \chi_\alpha 
\Big\}
\label{Sintmix}
\EN
where 
\EQA
L_{1\alpha}=(\gamma^0\theta_-)_\alpha, \quad
L_{1\alpha}^\dagger =-(\theta_-^\dagger\gamma^0)_\alpha, \nonumber\\
L_{2\alpha}=-(C^*\theta_-^{\dagger})_\alpha, \quad
L_{2\alpha}^\dagger =(\theta_- C)_\alpha. \quad
\EQN
The quadratic part of the action 
is given by $S^{\mbox{\scriptsize (quad)}}=S^{\mbox{\scriptsize (free)}}
+S^{\mbox{\scriptsize (int)}}_{\mbox{\scriptsize bos}}
+S^{\mbox{\scriptsize (int)}}_{\mbox{\scriptsize fermi}}
+S^{\mbox{\scriptsize (int)}}_{\mbox{\scriptsize mix}}$,
and the effective action is derived from the general 
expression (\ref{Gmix}). Note that $K_{{\mbox{\scriptsize bos}}}$
(or $K_{{\mbox{\scriptsize fermi}}}$) is read from (\ref{Sfree})
and (\ref{Sintbos}) (or from (\ref{Sfree})
and (\ref{Sintfermi})). Also, $K_{{\mbox{\scriptsize mix}}}$
is read from (\ref{Sintmix}).

We treat the vertices as an expansion around a reference 
time $\tau$ and obtain the effective action as
a double expansion in the number of vertices 
and the number of derivatives, following the procedure
described in ref.\cite{AS}.
Explicit form of the two-fermion (two $L$) terms of the 
effective action is given as%
\footnote{see ref.\cite{AS} for the terms which do not
contain fermions and  
for more details on the derivation.}
\EQA
\Gamma_{\theta^2} &=& \sum_{m,n=0}^{\infty}\sum_{D_i=0}^{\infty}
\left(\prod_{i=2}^{n+m+2} \!{1\over D_i!} \right) 
 (-1)^{(n+m)} {2\over \lambda^{m+2n+2}}
\nonumber\\
&& \times
\int\! d\tau\,
\Tr\, \left(V_{I_1,I_2} V^{(D_2)}_{I_2,I_3}\cdots 
V^{(D_n)}_{I_n, I_{n+1}}L^{\dagger(D_{n+m+1})}_{I_{n+1},\alpha}
\hX_{a_1}^{(D_{n+1})}\cdots\hX_{a_m}^{(D_{n+m})}
L_{I_1,\beta}^{(D_{m+n+2})} \right)\nonumber\\
&& \times 
\int 
{dk\over 2\pi} \left[
{1\over k^2+r'^2}(i\partial_k)^{D_2}\Big\{ {1\over k^2 +r'^2}
\cdots (i\partial_k)^{D_{n+m+1}}\Big\{
\frac{\rsl+ik}{k^2+r'^2}\tg^{a_1} (i\partial_k)^{D_{n+1}}\!
\Big\{ \right.\nonumber\\
&&\left. \cdots \tg^{a_m}(i\partial_k)^{D_{n+m}}
\Big\{ \frac{\rsl+ik}{k^2+r'^2}
(i\partial_k)^{D_{n+m+2}}{1\over k^2+r'^2}\Big\}\Big\} \Big\} \Big\}
 \right]_{\alpha\beta}
\EQN
For $D_i =0$, this reduces to
\EQA
\Gamma_{\theta^2}^{(d=0)} &=& \sum_{m,n=0}^{\infty}
 (-1)^{(n+m)} {2\over \lambda^{m+2n+2}}
\int\! d\tau\,
\Tr\, \left(V_{I_1,I_2} \cdots V_{I_n, I_{n+1}}L^{\dagger}_{I_{n+1},\alpha}
\hX_{a_1}\cdots\hX_{a_m}L_{I_1,\beta} \right)
\nonumber\\
&& \hspace*{-7mm}
\times
\int 
{dk\over 2\pi} \left[
{1\over (k^2+r'^2)^n}
\left(\frac{\rsl+ik}{k^2+r'^2}\tg^{a_1}\frac{\rsl+ik}{k^2+r'^2} 
\cdots \tg^{a_m} \frac{\rsl+ik}{k^2+r'^2} \right)_{\alpha\beta}
{1\over k^2+r'^2} \right].
\EQN
In the above equations, we have assumed $\hX_0=0$. The dependence
of the effective action on $\hX_0$  can be recovered by
replacing $\partial_\tau$ with the covariant derivative
$\partial_\tau+i[\hX_0,\:\:]$, or can be directly calculated
with a slight modification of the above prescription due to
the presence of the derivative
interaction $v_I \hX_0\partial_\tau v_I$.

\subsection{Fermionic terms of the one-loop effective action}


Following the method explained above, we calculate the
part of the effective action containing fermionic backgrounds.
We describe the $\theta^2$ terms in detail in section 3.2.1,
and briefly discuss the terms with more $\theta$'s in
section 3.2.2. The results are presented in Euclidean signature.
The rule for obtaining the Minkowskian effective action
from $\Gamma$ given below is to replace $\int\/ d\tau$ 
with $\int\/ dt$, and $D_\tau$ with $-i D_0$.

Interaction starts at $1/r^3$ order.
Note that only $\thm$ (but not $\thp$) appear 
non-trivially in the
effective action, for only $\thm$ couple to 
the quantum fields $v$ or $\chi$
as we have seen in section 2. 

\subsubsection{$\theta^2$ terms}

We summarize the result of $\theta^2$ terms of the effective action
$\Gamma_{\theta^2}(\hX_i^N,d)$
up to $N+d<4$,
where $N$ and $d$ are the numbers of 
$\hX_i$'s and derivatives contained in $\Gamma$, respectively.
We present the result 
by assembling the terms with the
same property.


There are two sets of terms which can be regarded as
having similar structures as the $\thm$ part of the
classical action:
\bea
\Gamma_{\theta^2}(d\!=\!0)_{A}
&=&
-\frac{N_4}{4}\frac{1}{r^3}
\int \!d\tau \,\tr\, \big(
\theta_-^\dagger \tilde{\gamma}^a[\hX_a,\theta_-] 
-[\hX_a, \theta_-^\dagger] \tilde{\gamma}^a\theta_-
\big)
\label{ad0r3}
\\
&& +\frac{3N_4}{4}\frac{r_b}{r^5}
\int\! d\tau\, \str \,
\big(
\theta_-^\dagger \tilde{\gamma}^a[\hX_a,\theta_-] 
-[\hX_a, \theta_-^\dagger] \tilde{\gamma}^a\theta_-
\; ; \; \hX_b \big)
\\
&& +\frac{3N_4}{8}\frac{1}{r^5}
\int\! d\tau \,\str 
\, \big(
\theta_-^\dagger \tilde{\gamma}^a[\hX_a,\theta_-] 
-[\hX_a, \theta_-^\dagger] \tilde{\gamma}^a\theta_-
\, ; \, (\hX_b)^2 \big)
\label{ad0r51}
\\ 
&& -
\frac{15N_4}{8 }\frac{r_{b}r_{c}}{r^7}
\int \!d\tau \str \,
\big(
\theta_-^\dagger \tilde{\gamma}^a[\hX_a,\theta_-] 
\!-\![\hX_a, \theta_-^\dagger] \tilde{\gamma}^a\theta_-
\, ; \,\hX_b,\hX_c \big),
\eea
\bea
\Gamma_{\theta^2}(d\!=\!1)_{A}
&=& 
\frac{N_4\lambda}{4}\frac{1}{r^3}
\int\! d\tau \,\tr \,\big(
\theta_-^\dagger D_\tau\theta_- 
\!-\!D_\tau \theta_-^\dagger \theta_-
\big)
\label{ad1r3}
\\
&& -\frac{3N_4\lambda}{4}\frac{r_b}{r^5}
\int \!d\tau\, \str 
\, \big(
\theta_-^\dagger D_\tau\theta_- 
\!-\!D_\tau \theta_-^\dagger \theta_-
\, ; \, \hX_b \big)
\\
&& -\frac{3N_4\lambda}{8}\frac{1}{r^5}
\int \!d\tau \, \str \,
\big(\theta_-^\dagger D_\tau \theta_- 
\!-\!D_\tau \theta_-^\dagger \theta_-
\, ; \, (\hX_b)^2 \big)
\label{ad1r51}
\\ 
&& +
\frac{15N_4\lambda}{8 }\frac{r_{b}r_{c}}{r^7}
\int\! d\tau \,\str \,
\big(
\theta_-^\dagger D_\tau \theta_- 
\!-\!D_\tau \theta_-^\dagger \theta_-
\, ; \,\hX_b,\hX_c \big).
\eea
Here $\str(K_1\cdots K_m; y_1,y_2,\cdots, y_n)$
means that the trace operation is taken after
symmetrizing all $K_i$'s and $y_j$'s 
but keeping the location of $K_1$ and the order of $K_i$'s.
Note that each $K_i$ is 
$\hX_a$, $\hX_m$, $\thm$, or commutators (or covariant
derivatives) of them.
For example, 
\bea
\str \,
\big(
\thmd D_\tau \thm \, ; \,\hX_b,\hX_c 
\big)
&=&
\frac{1}{6}
\tr \,
\big(
\thmd D_\tau \thm \hX_b \hX_c + 
\thmd \hX_b \hX_c D_\tau \thm + 
\thmd \hX_b D_\tau \thm \hX_c  
\nonumber\\
&& \qquad +
\thmd D_\tau \thm \hX_c \hX_b +
\thmd  \hX_c \hX_b D_\tau \thm +
\thmd \hX_c D_\tau \thm \hX_b  
\big).
\nonumber
\eea
The terms in $\Gamma_{\theta^2}(d\!=\!0)_{A}$ and
$\Gamma_{\theta^2}(d\!=\!1)_{A}$ are given 
by the non-abelian Taylor expansion of the leading terms
(\ref{ad0r3}) and (\ref{ad1r3}) 
$$
{1\over r^3}\rightarrow \frac{1}{r^3}-\frac{3}{r^5}r_a\hX_a
+\frac{1}{2}
\left(
-3\frac{\delta_{ab}}{r^5} +15\frac{r_a r_b}{r^7} 
\right)\hX_a\hX_b
+\cdots ,
$$
and following the symmetrized
trace prescription \cite{TR1, My}, 
except for the fact that (\ref{ad0r51}) and
(\ref{ad1r51}) are of the form $\str (\;\ast\; ; (\hX_b)^2)$
rather than $\str (\;\ast\; ; \hX_b, \hX_b)$.
Further discussion on the ordering of matrices will be given
in the next section when we consider the supersymmetry of
the effective action.


Terms with multiple number of gamma matrices 
are given as follows:
\bea
\!\!\Gamma_{\theta^2}(d\!=\!0)_{B}
&=&
\frac{3 N_4r_a}{16\,r^5} 
\int \!d\tau \,  
\tr\,\big( {\thmd}\tilde{\gamma}^{aa_1a_2}[\hX_{a_1},\hX_{a_2}] \thm
+{\thmd} \tilde{\gamma}^{aa_1a_2} \thm [\hX_{a_1},\hX_{a_2}] 
\big)
\\
&&\hspace*{-2.2cm} +\,\frac{3 N_4}{16\,r^5} 
\int \!d\tau \,  
\tr \,\big( {\thmd}\tilde{\gamma}^{a_1a_2a_3}[\hX_{a_1},\hX_{a_2}] \hX_{a_3} \thm
+{\thmd} \tilde{\gamma}^{a_1a_2a_3}\thm[\hX_{a_1},\hX_{a_2}]\hX_{a_3} 
\big)
\label{nonstr1}
\\
&&  \hspace*{-2.2cm}
 -\,
\frac{15 N_4}{16}\frac{ r_ar_b}{r^7} 
\int \!d\tau \,  
\str\,\big( {\thmd}\tilde{\gamma}^{aa_1a_2}[\hX_{a_1},\hX_{a_2}] \thm
+{\thmd} \tilde{\gamma}^{aa_1a_2}\thm[\hX_{a_1},\hX_{a_2}] \;; \; \hX_b
\big),
\eea
\bea
\Gamma_{\theta^2}(d\!=\!1)_{B}
&=& -
\frac{3 N_4\lambda}{8}\frac{r_b}{r^5} 
\int \!d\tau \,  
\tr\,\big( \thmd\tg^{ab} D_\tau \hX_a \thm +\thmd\tg^{ab} \thm D_\tau  \hX_a \big)
\\
&& -
\frac{ 3N_4\lambda}{16}\frac{1}{r^5} 
\int \!d\tau \,  
\tr \,\big( \thmd \tg^{ab} \hX_b D_{\tau} \hX_a \thm 
+ \thmd \tg^{ab}  D_{\tau} \hX_a \hX_b \thm
\nonumber\\ 
&& \hspace*{3cm}
+ \thmd \tg^{ab} \thm \hX_b D_{\tau} \hX_a
+ \,\thmd \tg^{ab} \thm D_{\tau} \hX_a \hX_b 
\big)
\label{nonstr2}
\\
&&
+
\frac{15 N_4\lambda}{8}\frac{r_br_c}{r^7} 
\int \!d\tau \,  
\str \,\big( \thmd \tg^{ac} D_\tau \hX_a \thm + \thmd \tg^{ac} \thm D_\tau \hX_a \,;\,
\hX_b \big) .
\eea

In addition, there are commutator corrections to
$\Gamma_{\theta^2}(d\!=\!0,1)_{A,B}$:
\bea
\Gamma_{\theta^2}(d\!=\!0)_{C}
\!&\!=\!&\!
\frac{3N_4}{16r^5}
 \int \!d\tau \,  \tr \big(
{\thmd} \tilde{\gamma}^{b} \big[\hX_a, \big[\hX_{a},\hX_{b} \big] \big] \thm
\!-\!{\thmd}\tilde{\gamma}^{b}\thm \big[\hX_a, \big[\hX_{a},\hX_{b} \big] \big] 
\big)
\\
\Gamma_{\theta^2}(d\!=\!1)_{C}
&\!=\!& \frac{3N_4\lambda}{16r^5}
\int \!d\tau \, \tr \big(
\thmd \big[\big[\hX_a,\thm\big],D_\tau \hX_a\big]
+ \big[\big[\thmd, \hX_a \big] , D_\tau \hX_a\big] \thm  
\big)
\\ 
\!\!&\!+\!\!&\! \frac{N_4\lambda}{32r^5}
\int \!\!d\tau \, \tr \big(
D_\tau \thmd \tg^{ab} \big[ \big[\hX_a, \hX_b \big], \thm \big]
\!-\! \big[\thmd, \big[\hX_a, \hX_b \big] \big] \tg^{ab} D_\tau\thm
\big)
\eea 

Terms containing $\hX_m$ are collected as:
\bea
\Gamma_{\theta^2}(d\!=\!0)_{\phi}
&=& 
-\frac{3N_4r_a}{16\,r^5} \int \!d\tau \,
\tr\,\big(\Thmd[\hX_m, \hX_n]\Gamma^{0a}\Gamma^{mn}\Thm \big)
\label{d0Xmr4}
\\
&&
-\frac{3N_4}{16\,r^5} \int \!d\tau \,
\tr\, \big(\Thmd[\hX_m, \hX_n]\Gamma^{0a}\Gamma^{mn}\Thm \hX_a
\big)
\label{d0Xmr51}
\\
&&+\frac{15N_4}{16}\frac{r_ar_b}{r^7}\int \!d\tau \, 
\str\,
\big(\Thmd[\hX_m, \hX_n]\Gamma^{0a}\Gamma^{mn}\Thm\,;\,\hX_b 
\big),
\\
\Gamma_{\theta^2}(d\!=\!1)_{\phi}
\!&\!=\!&\!\!
\frac{N_4\lambda}{16\,r^5} \!
\int \!d\tau \,  
\tr\, \big( D_\tau{\Thmd}[\hX_m, \hX_n]\Gamma^{mn}\Thm
\!-\!{\Thmd}[\hX_m, \hX_n]\Gamma^{mn}D_{\tau}{\Thm}
\big).
\label{d1Xmr5}
\eea
Note that we have used the notation of 10D gamma matrices 
and spinors to write (\ref{d0Xmr4})-(\ref{d1Xmr5}).
The expressions in 6D notation is given by substituting
the particular representation (\ref{10Dgamma}) and (\ref{4Dgamma})
of 10D gamma matrices
and the parametrization (\ref{Theta10D6D}) 
of Majorana-Weyl spinor $\Theta$ 
setting $\thp=0$. For example, (\ref{d0Xmr4}) is
written in 6D notation as
\EQAN
&&+\frac{3N_4r_a}{16\,r^5} \int \!d\tau \,
\tr\,
\Big\{\thmd 
\tg^a \thm ([\phi_1,\bar{\phi}_1]+[\phi_2,\bar{\phi}_2])
+\thmd ([\phi_1,\bar{\phi}_1]+[\phi_2,\bar{\phi}_2])
\tg^a \thm\\
&&\hspace*{3.5cm} -2\thm B[\phi_1,\phi_2] \tg^a \thm
-2\thmd [\bar{\phi}_1,\bar{\phi}_2] \tg^a B^* \thmd \Big\}.
\EQNN
Terms in $\Gamma_{\theta^2}(d\!=\!0)_{B}$,
$\Gamma_{\theta^2}(d\!=\!1)_{B}$
and $\Gamma_{\theta^2}(d\!=\!0)_{\phi}$
 have the structure of non-abelian 
Taylor expansion of $r_a/r^5$:
$$
{r_a\over r^5}\rightarrow \frac{r_a}{r^5}
+ \left(
\frac{\delta_{ab}}{r^5} -5\frac{r_a r_b}{r^7}
\right) \hX_b +\cdots .
$$
They obey symmetrized trace prescription 
except for $1/r^5$-terms (\ref{nonstr1}), (\ref{nonstr2})
and (\ref{d0Xmr51}), as in the case
of $\Gamma_{\theta^2}(d=0,1)_A $.

There also exist terms with two derivatives:
\begin{equation}
\Gamma_{\theta^2}(d\!=\!2)
=
\frac{N_4 \lambda^2}{8r^5}
\int \!d\tau \, \tr 
\big( D_\tau \thmd \tg^a \big[ \hX_a, D_\tau\thm\big] -
2 \thmd \tg^a \big[D_\tau^2 \hX_a, \thm \big] \big).
\end{equation}
This completes the analysis of $\Gamma_{\theta^2}(\hX_i^N,d)$
for $N+d<4$.


To summarize, we obtained 
\bea
\Gamma_{\theta^2}
&=& \Gamma_{\theta^2}(d\!=\!0,1)_{A}+
\Gamma_{\theta^2}(d\!=\!0,1)_{B}
+\Gamma_{\theta^2}(d\!=\!0,1)_{\phi}
\nonumber\\
&&\;+\Gamma_{\theta^2}(d\!=\!0,1)_{C}
+\Gamma_{\theta^2}(d\!=\!2)\quad +{\cal O}(D_\tau^d \hX^N\theta^2 ;N\!+\!d\ge 4)
\eea
for two-fermion part of the one-loop effective action.

\subsubsection{$\theta^{2n}$ terms}

The leading terms of $\Gamma_{\theta^{2n}}$
are given by 
\bea
\Gamma_{\theta^{2n}}((\hX)^0,d\!=\!0)
&=&N_4
\left(\frac{2}{\lambda^2}\right)^n \frac{1}{n}
\int \!d\tau\, 
\tr(L^\dagger_{I_n \alpha_1} L_{I_1 \beta_1} L^\dagger_{I_1 \alpha_2} L_{I_2 \beta_2} 
\cdots L^\dagger_{I_{n-1} \alpha_n} L^\dagger_{I_n \beta_n} )
\nonumber\\
&& \times \int\frac{dk}{2\pi} \left[
\left(\frac{\rsl+ik}{k^2+r^2}\right)_{\alpha_1\beta_1}
\left(\frac{\rsl+ik}{k^2+r^2}\right)_{\alpha_2\beta_2}
\cdots
\left(\frac{\rsl+ik}{k^2+r^2}\right)_{\alpha_n\beta_n}
\right]
\nonumber\\
&\propto& 
\frac{N_4\lambda^{n-1}}{r^{3n-1}}
\int\! d\tau\, \tr (\thm^{2n}).
\eea
For example, four-fermion terms without $X_i$ insertions
are given as
\bea
&& \hspace*{-1.3cm}\Gamma_{\theta^4}((\hX)^0,d\!=\!0)
\nonumber\\
&=& 
- \frac{N_4\lambda}{16}\frac{1}{r^5} 
\int\! d\tau\, \tr 
\left[(\thmd \thm) (\thmd \thm) + (\thm \thmd)( \thm \thmd) 
-2(\thm B \thm)(\thmd B^*\thmd ) \right]
\nonumber\\ 
&& 
+\frac{5N_4 \lambda}{16}\frac{r_a r_b}{r^7} 
\int\! d\tau\, \tr \Big[(\thmd\tg_a \thm) (\thmd \tg_b\thm) + 
(\thm\tg_a^* \thmd)( \thm \tg_b^*\thmd) 
\nonumber\\
&& \hspace*{7cm} 
-2(\thm B \tg_a \thm)(\thmd \tg_bB^*\thmd ) \Big].
\eea
This does not vanish unlike the case of $\Gamma_{\theta^2}$.



\section{Supersymmetry of the effective action}

In this section, we discuss the supersymmetry of the 
effective action. 
The one-loop effective action which was obtained 
by integrating
out the 0-4 sector fields satisfies the Ward identity
corresponding to the SUSY invariance of the classical action 
of BD Matrix theory.
As we show in Appendix B, at the one-loop level,
the Ward identity can be written in the form 
\EQ
\delta^{(1)} S_0+\delta^{(0)} S^{(1)}=0
\label{1loopSUSY}
\EN
which states the invariance
of the effective action under an effective SUSY transformation.

Let us explain the meaning of (\ref{1loopSUSY}).
The tree action $S_0$ is the part of BD action containing
only the 0-0 sector fields which is given in (\ref{S0})
and is the same as the BFSS action.
It is invariant under the SUSY transformation which does not
involve 0-4 sector fields $\delta^{(0)}S_0=0$. 
The transformation law for $\delta^{(0)}$, which we call
the tree-level SUSY transformation is given by 
(\ref{BDSUSY00}) with a replacement 
$X_a \rightarrow \hX_a$.
Remember that we have divided
the background  $X_a$ as $X_a=r_a+\hX_a$ in (\ref{Xdecomp}).
In our construction, only $\hX_a$ is transformed and 
$r_a$ is kept {\it fixed.}\footnote{This assignment
is possible without losing generality. Note that
we have not imposed the tracelessness for $\hX_a$.} 
The discussion in Appendix B shows that
$\delta^{(1)}$ is given simply by
the one-loop expectation value of the SUSY transformation 
for the classical action.
Since only $\delta\thm$ has the part containing
the quantum fields
($\delta'\thm$ defined in (\ref{BDSUSYthmp})), 
only $\delta^{(1)}\thm$ is non-zero and is given by
\EQA
\delta^{(1)}\thm& =& \langle \delta'\thm \rangle\nonumber\\
&=&\frac{i}{\lambda} 
\left[ \big(- \big\langle v_1v_1^\dagger \big\rangle 
+ \big\langle v_2v_2^\dagger \big\rangle \big) \eta_- 
-2 \big\langle v_2v_1^\dagger \big\rangle B^* \eta^\dagger
\right].
\label{del1thm}
\EQN


\subsection{Explicit forms of the one-loop corrected
SUSY transformation}


We give the 
one-loop corrected SUSY transformation of $\thm$
by explicitly calculating (\ref{del1thm}).
Results at order $1/r^3$ and $1/r^4$
are given respectively as
\begin{equation}
{1\over g_sl_s} \delta^{(1)} \thm \big|_{1/r^3} =
 i \frac{N_4}{4r^3} 
\left\{ \big( [\phi_1,\bar{\phi}_1]+[\phi_2,\bar{\phi}_2]
\big) \eta 
-2 [\bar{\phi}_1,\bar{\phi}_2]B^* \eta^\dagger
\right\},
\label{d1a0}
\end{equation}
\bea
\frac{1}{g_sl_s} 
\delta^{(1)} \thm \big|_{1/r^4} &=&
- i \,\frac{3N_4}{4}\frac{r_a}{r^5}
\Big\{ \sym\big( [\phi_1,\bar{\phi}_1]+[\phi_2,\bar{\phi}_2]
\, ; \, \hX_a \big) \eta
\nonumber\\
&&\hspace*{5cm} 
-2 \, \sym \big( [\bar{\phi}_1,\bar{\phi}_2] \,;\, \hX_a\big) B^* \eta^\dagger
\Big\}
\label{d1a1}
\\
&& +\,i \,\frac{3N_4\lambda}{8}\frac{r_a}{r^5}
\Big\{ \big( \thmd \tg^a  \thm -\thm \tg^{a*}\thmd
 \big) \eta 
-2  \big( \thm B \tg^a  \thm
\big) B^* \eta^\dagger
\Big\}.
\label{d1b0}
\eea
Here $\sym( K_1\cdots K_m; y_1, \cdots, y_n)$
is the symmetrization of all $K_i$'s and $y_j$'s 
without changing the order of $K_i$'s.
Note that there is a relation
$$\str(K_1 K_2\cdots K_m  ; \,\ast\,)= 
\tr\,( K_1\sym(K_2\cdots K_m; \,\ast\,)).$$

For order $1/r^5$, we divide the result 
into two parts 
$
 \delta^{(1)}\thm \big|_{1/r^5}= \delta^{(1)}\thm \big|_{1/r^5}^{(a)} 
+\delta^{(1)}\thm \big|_{1/r^5}^{(b)}  
$:
\bea
&& \hspace*{-8mm} \frac{1}{g_sl_s} \delta^{(1)} \thm \big|_{1/r^5}^{(a)} 
\nonumber\\
&& \hspace*{-6mm}=
- i \,\frac{3N_4}{8\,r^5}
\Big\{ \sym\big( [\phi_1,\bar{\phi}_1]+[\phi_2,\bar{\phi}_2]
\, ; \, (\hX_a)^2 \big) \eta
-2 \, \sym \big( [\bar{\phi}_1,\bar{\phi}_2] \,;\, (\hX_a)^2 \big) B^* \eta^\dagger
\Big\}
\label{d1a21}
\\
&& \hspace*{-6mm}
+i\, \frac{15N_4}{8}\frac{r_ar_b}{r^7}
\Big\{ \sym\big( [\phi_1,\bar{\phi}_1]\!+\![\phi_2,\bar{\phi}_2]
\, ;  \hX_a,\hX_b \big) \eta
\!-\!2 \, \sym \big( [\bar{\phi}_1,\bar{\phi}_2] \,; \hX_a,\hX_b \big) B^* \eta^\dagger
\Big\}
\label{d1a22}
\\
&&\hspace*{-6mm}
+ i\, \frac{3N_4\lambda}{8\,r^5}
\Big\{ \big( \thmd \tg^b \hX_b \thm \! -\thm \hX_b \tg^{b*} \thmd
 \big) \eta \!
-2  \big( \thm B \tg^b \hX_b \thm
\big) B^* \eta^\dagger
\Big\}
\label{d1b11}
\\
&&\hspace*{-6mm}
- i \,\frac{15N_4\lambda}{8}\frac{r_ar_b}{r^7}
\Big\{ \sym\big( \thmd \tg^a \thm \! -\thm \tg^{a*} \thmd
\,;\hX_b \big) \eta \!
-2 \, \sym\big( \thm B \tg^a \thm
\, ; \hX_b \big) B^* \eta^\dagger
\Big\}
\label{d1b12}
\eea
and
\bea
&& \hspace*{-15mm} \frac{1}{g_sl_s} \delta^{(1)} \thm \big|_{1/r^5}^{(b)} 
\nonumber\\
&& \hspace*{-6mm}=
 i \,\frac{3N_4}{16\,r^5}
\Big\{ \big( \big[[\phi_1,\phi_2],[\bar{\phi}_1,\bar{\phi}_2]\big]
\big) \eta 
+ \,\big[[\bar{\phi}_1,\bar{\phi}_2],
\big([\phi_1,\bar{\phi}_1]+[\phi_2,\bar{\phi}_2]\big)
\big]B^* \eta^\dagger
\Big\}
\\
&& \hspace*{-2mm}
- i \,\frac{N_4\lambda^2}{16\,r^5}
\Big\{ D_0^2 \big( [\phi_1,\bar{\phi}_1]+[\phi_2,\bar{\phi}_2]
\big) \eta 
-2 \,D_0^2\big([\bar{\phi}_1,\bar{\phi}_2]\big) B^* \eta^\dagger
\Big\}
\\
&& \hspace*{-2mm}
- \, \frac{N_4\lambda^2}{4\,r^5}
\Big\{ \big( \thmd D_0 \thm - D_0 \thm \thmd
\big) \eta 
-2\, \big(\thm B D_0 \thm \big)
B^* \eta^\dagger
\Big\}.
\eea
Note that (\ref{d1a0}), (\ref{d1a1}), (\ref{d1a21}) and (\ref{d1a22})
correspond to the first three terms of 
non-abelian Taylor expansion of 
$1/r^3$ around $r_a$.
Also, (\ref{d1b0}), (\ref{d1b11}) and (\ref{d1b12})
correspond to the expansion of $r_a/r^5$. 


Our corrected SUSY transformation closes to the translation
plus a field dependent gauge transformation \cite{WF}.
By using the equations of motion for the effective action
$S_0+S^{(1)}$, we have shown 
$$
[ \delta^{(0)}_{\epsilon}+\delta^{(1)}_{\epsilon}, 
\delta^{(0)}_{\eta}+\delta^{(1)}_{\eta}] 
 \,\phi \, \Big|_{1/r^4}
\sim (D_M \phi\, \bar{\epsilon}\Gamma^M\eta)+
(\mbox{EOM for $\phi$}) + {\cal O}\big((g_sl_s)^2\big)
$$
up to the order $1/r^4$ for any field $\phi$
in $S_0+S^{(1)}$.


\subsection{On the structure of the supersymmetric action}

In this subsection, we try to clarify the structure
of the supersymmetric action which we have obtained, 
by identifying  the blocks which transform 
within themselves.
Firstly, since $r_a$ is kept fixed in the 
effective SUSY transformation,
the invariance of the one-loop effective action 
(\ref{1loopSUSY}) holds at each order of $1/r$.
In addition, the effective action at order $1/r^5$
is further divided into two invariant blocks.

We shall examine the $\theta^0$- and  $\theta^2$-terms at each order
of $1/r$%
\footnote{Note that $\Gamma$ which are discussed in this section 
denote the parts of the Minkowskian effective action $S^{(1)}$.}. 
We include the bosonic terms obtained in the
previous paper~\cite{AS}.
Order $1/r^3$ terms of the effective action 
\EQA
\Gamma_{1/r^3}&=&
{\lambda\over r^3}
 \int dt \Tr \Big\{ {1\over 4}D_0\hX_a D_0\hX_a
+{1\over 8\lambda^2}[\hX_a,\hX_b][\hX_a,\hX_b]
-{1\over 4\lambda^2}[\phi_1,\phi_2][\bar{\phi}_1,\bar{\phi}_2]
\nonumber\\
&&\quad +{1\over 16\lambda^2}([\phi_1,\bar{\phi}_1]
+[\phi_2,\bar{\phi}_2])^2
-{i\over 2}\thmd D_0 \thm -{1\over 2\lambda}\thmd
\tg^a [\hX_a,\thm]\Big\}.
\label{orderr3}
\EQN
satisfy (\ref{1loopSUSY}) along with (\ref{d1a0}).

Order $1/r^4$ terms of the effective action
\EQA
\Gamma_{1/r^4}&\!=\!&-3 N_4 \lambda {r_a\over r^5} \int dt 
\bigg\{
\str \bigg( {1\over 4}D_0\hX_c D_0\hX_c 
+{1\over 8\lambda^2}[\hX_c,\hX_d][\hX_c,\hX_d]
\nonumber\\
&&\qquad \hspace*{3cm}
-{1\over 4\lambda^2}[\phi_1,\phi_2][\bar{\phi}_1,\bar{\phi}_2]
+{1\over 16\lambda^2}([\phi_1,\bar{\phi}_1]
+[\phi_2,\bar{\phi}_2])^2
\, ;\hX_a \bigg) 
\nonumber\\
&&\qquad
+{i\over16\lambda}\epsilon_{a a_1a_2a_3a_4}
\str \big( [\hX_{a_2},\hX_{a_3}]D_0 \hX_{a_4}\,; \hX_{a_1}\big)
\nonumber\\ 
&&\qquad +\,\str \bigg(\!-{i\over 4}( \thmd D_0 \thm \!-\!D_0 \thmd \thm) 
-{1\over 4\lambda} (\thmd \tg^c [\hX_c,\thm]
\!-\![\hX_c, \thmd]\tg^c \thm) \,;\hX_a \bigg)
\nonumber\\
&&\qquad + \, \tr\bigg({i\over 8}(D_0\hX_b\thmd \tg^{ab}\thm
+\thmd D_0\hX_b\tg^{ab}\thm)
\nonumber\\
&&\qquad \qquad
-{1\over 16\lambda} ([\hX_b,\hX_c]\thmd \tg^{abc}\thm
+\thmd [\hX_b,\hX_c] \tg^{abc}\thm)\nonumber\\
&&\qquad \qquad -{1\over 16\lambda} \{ ([\phi_1,\bar{\phi}_1]
+[\phi_2,\bar{\phi}_2])\thmd\tg^a\thm
+\thmd([\phi_1,\bar{\phi}_1]
+[\phi_2,\bar{\phi}_2])\tg^a\thm\nonumber\\
&&\qquad \hspace*{4cm}
-2 [\phi_1,\phi_2] \thm B \tg^a \thm
-2 [\bar{\phi}_1,\bar{\phi}_2] \thmd B^* \tg^a \thmd\}
\bigg)\bigg\}
\label{orderr4}
\EQN
satisfy (\ref{1loopSUSY}) along with (\ref{d1a1}), (\ref{d1b0}).

By examining $\delta^{(0)}S^{(1)}$, we find that the following 
terms of the $1/r^5$ effective action transform among themselves
(with a certain class of terms in $\delta^{(1)}S_0$).
\EQA
\Gamma_{1/r^5}^{{\rm (I)}}
&=& N_4 \lambda
\left( {15 r_a r_b\over 2r^7}
-{3 \delta_{ab}\over 2 r^5}\right) \int dt 
\bigg\{
\str \bigg( {1\over 4}D_0\hX_c D_0\hX_c 
+{1\over 8\lambda^2}[\hX_c,\hX_d][\hX_c,\hX_d]
\nonumber\\
&&\qquad \hspace*{3cm}
-{1\over 4\lambda^2}[\phi_1,\phi_2][\bar{\phi}_1,\bar{\phi}_2]
+{1\over 16\lambda^2}([\phi_1,\bar{\phi}_1]
+[\phi_2,\bar{\phi}_2])^2
\,;\hX_a,\hX_b \bigg) 
\nonumber\\
&& \quad
+{i\over10\lambda}\epsilon_{a a_1a_2a_3a_4}
\, \str \big( [\hX_{a_2},\hX_{a_3}]D_0 \hX_{a_4}\,; \hX_{a_1},\hX_b\big)
\nonumber\\ 
&&\quad +\,\str \bigg(\!
-{i\over 4}( \thmd D_0 \thm \!-\!D_0 \thmd \thm) 
-{1\over 4\lambda} (\thmd \tg^c [\hX_c,\thm]
\!-\![\hX_c, \thmd]\tg^c \thm) \,;\hX_a,\hX_b \bigg)
\nonumber\\
&& \quad +{i\over 4}\, \str \big( D_0\hX_c\thmd \tg^{ac}\thm
+\thmd D_0\hX_c\tg^{ac}\thm \,;\hX_b \big)
\nonumber\\
&& \quad -{1\over 8 \lambda} \, \str \big([\hX_{a_1},\hX_{a_2}]\thmd 
\tg^{aa_1a_2}\thm +\thmd [\hX_{a_1},\hX_{a_2}] 
\tg^{aa_1a_2}\thm \,;\hX_b \big)
\nonumber\\
&& \quad -{1\over 8\lambda} \,\str \Big( ([\phi_1,\bar{\phi}_1]
+[\phi_2,\bar{\phi}_2])\thmd\tg^a\thm
+\thmd([\phi_1,\bar{\phi}_1]
+[\phi_2,\bar{\phi}_2])\tg^a\thm\nonumber\\
&& \quad \hspace*{3.5cm} -2 [\phi_1,\phi_2] \thm B \tg^a \thm
-2 [\bar{\phi}_1,\bar{\phi}_2] \thmd B^* \tg^a \thmd 
\,; \hX_b
\Big) \bigg\}
\label{orderr5I}
\EQN



The other block $\Gamma_{1/r^5}^{{\rm(II)}}$ consists of 
the rest of the terms of $1/r^5$ effective action. 
They are written with extra number of commutators 
compared to (\ref{orderr5I}). We do not present all the
terms explicitly,
but give an example. The part which contain six $\hX_a$'s (and
no derivatives or fermions) takes the form
\EQA
\Gamma_{1/r^5}^{{\rm (II)}}\big|_{\hX_a^6}
&=&-{3N_4 \over 2 r^5}\int\! dt\, \Big\{
-{1\over 24\lambda} \tr ( [\hX_a,[\hX_b,\hX_c]]
[\hX_a,[\hX_b,\hX_c]])\\
&& +{1\over 12\lambda} \tr
( [\hX_b,[\hX_a,\hX_c]] [\hX_a,[\hX_b,\hX_c]])
-{1\over 24\lambda} \tr ( [\hX_a,[\hX_a,\hX_c]]
[\hX_b,[\hX_b,\hX_c]])\Big\}\nonumber.
\label{orderr5II}
\EQN

Criterion for discriminating $\Gamma_{1/r^5}^{{\rm (I)}}$ and
$\Gamma_{1/r^5}^{{\rm (II)}}$ is given by the following number
associated with each term:
\EQ
n=d+{1\over 2}n_{\theta}+n_c.
\label{order}
\EN
Here, $d$ is the number of derivatives, $n_\theta$ is the
number of fermions and $n_c$ is the number of commutators
{\it in the symmetrized trace}. (Symmetrization is applied 
regarding the commutator as a single unit.)
Under $\delta^{(0)}S^{(1)}$, the number $n$ is preserved 
(uniformly increase by 1/2) 
as we see from the transformation law (\ref{BDSUSY00}),
(\ref{BDSUSYthm0}). Also, terms in $\delta^{(1)}S_0$ are
classified in the same way.
Thus, terms which are connected by SUSY transformation must
have the same $n$.
The first two terms of the RHS of (\ref{order})
is called the `order' \cite{H} and used to specify
the SUSY invariants in the case of abelian v.e.v. 
Our discussion here is the non-abelian generalization
of that concept.
Terms in $\Gamma_{1/r^3}$,
$\Gamma_{1/r4}$ and $\Gamma_{1/r^5}^{{\rm (I)}}$
have $n=2$, and terms in $\Gamma_{1/r^5}^{{\rm (II)}}$
have $n=4$.

We note here that the 
bosonic terms in $\Gamma_{1/r^3}$,
$\Gamma_{1/r4}$ and $\Gamma_{1/r^5}^{{\rm (I)}}$
are the ones which result from the Taylor and Van Raamsdonk's
proposal when we apply it to L5-brane background.
We can see that the proposed currents which couple to SUGRA
fields have $n=2$, from the explicit forms in ref.\cite{TR1, AS}.
Bosonic terms of the matrix expressions for multipole moments are 
obtained from the
currents by inserting $\hX_i$ with symmetric ordering, thus
also have $n=2$. For a detailed description of the Taylor and
Van Raamsdonk's proposal applied to the present background,
see our previous paper \cite{AS}.

We also note that the fermionic terms in 
$\Gamma_{1/r^3}$,
$\Gamma_{1/r4}$ and $\Gamma_{1/r^5}^{{\rm (I)}}$
are not of the same forms as the ones arising from the 
above proposal, which is based on the analysis of BFSS model. 
Whether the two forms of the fermionic terms are physically 
equivalent or not is not clear at present.

\section{Discussion}
We have calculated the one-loop effective action of 
Berkooz-Douglas Matrix theory by integrating out the 
0-4 sector fields. Extending the analysis of ref.\cite{AS},
we obtained the fermionic part of the effective action of
multiple D0-branes in the longitudinal 5-brane background.
As we have seen in section 4, the action is manifestly
invariant under an effective SUSY transformation,
which is the consequence of the SUSY invariance of the
classical action of BD Matrix theory.
The transformation law is given by the 
expectation value of the SUSY transformation
of classical action evaluated at one-loop order.
We have studied the transformation law and identified
the structure which are closed under the effective
SUSY transformation.

Primary motivation of our study
is to understand the dynamics
and the structure of symmetries of this matrix model. 
In addition, 
we regard our analysis as a first step towards constructing
the supersymmetric action of multiple D-branes in the 
supergravity backgrounds. We should note that
from the 10D perspective,
our model is nothing but the SYM which takes into account 
only the lowest modes of open strings.
Thus, the effective action is guaranteed to be valid only
when the distance between D0-branes and D4-branes is short.
However, some of the terms in the effective action continues
to be valid 
when the separation becomes large. It is well known that
for the $(\partial_t\hX{}_i)^2$-term in the abelian case, the SYM 
effective action gives exact results, due to the cancellation
of the open-string higher modes \cite{DKPS}. 
It may be the case that 
same kind of non-renormalization properties
exist for all the other $n=2$ terms which we 
identified in section 4.2. 
Further analysis of the terms in this class
with higher orders in $\theta$ or $1/r$ 
should be important in this regard.


Of course, to determine which terms of the SYM effective 
action is valid for large separation, we must study
the cylinder amplitudes between the branes, or prove
non-renormalization theorems in SYM. For the case of
multiple branes, both of them are not well-understood at present.
Those are important subjects for future studies.


As mentioned in the introduction, $\kappa$-symmetric Born-Infeld
action for
a single D-brane can be written in arbitrary background, in 
principle.
We will be able to understand the structure
of the supersymmetric action further by comparing the
abelian part of our effective action with the gauge-fixed
Born-Infeld 
action of D0-branes in D4-brane background. Especially,
it will be interesting if the difference  
between the fermionic part in 
the effective action of BD Matrix theory and the 
ones present in the proposal based on the analysis of
BFSS model is explained by a difference in the
gauge choice for $\kappa$-symmetry.

Two-loop study of BD Matrix model should give important
implications for the  connection between
matrix models and gravity. In the first place, we must clarify
to what extent the Matrix theory reproduces supergravity
result in the case of abelian matrix background.
Two-loop effective action will also shed light on the coupling 
of multiple D-branes to non-weak background fields.

Finally, it is an interesting problem to find  classical 
solutions of the effective action which preserve SUSY. 
As argued by Myers \cite{My}, background 4-form field strength 
produced by D4-branes should give rise to 
non-abelian configuration of D0-branes.
It will be possible to discuss the BPS
property of those configurations
from the effective action which we have obtained. 

\vspace{0.4cm}
\paragraph{Acknowledgements}

We would like to thank Y. Kazama, T. Muramatsu, N. Sakai and T. Yoneya
for discussions and comments.

\appendix
\renewcommand{\theequation}{\Alph{section}.\arabic{equation}}

\section{Convention for the representations of spinors and
gamma matrices}
In this paper, we mainly use the complex 6D Weyl spinors.
We summarize here
our conventions for  spinors and gamma matrices.
The fermions from the 0-0 sector are given by
Majorana-Weyl spinors in 10D.  We explain that the fermionic
terms in $S_0$ in (\ref{S0}) is given from the
familiar 10D notation by choosing an 
explicit representation.


We consider fermionic matrix field $\Theta$ which satisfy
the  10D Majorana-Weyl condition
\begin{equation}
\bar{\Gamma}{}^{(10)}\Theta=\Theta,\qquad
\Theta^\dagger =B^{(10)} \Theta
\label{MWcond}
\end{equation}
where $\bar{\Gamma}^{(10)}=\Gamma^0\Gamma^1\cdots\Gamma^9$
is the 10D chirality matrix, and $B^{(10)}$ is defined by
\[
B^{(10)}\Gamma^M B^{(10)}{}^{-1}=-\Gamma^M{}^* \qquad (M=0,\ldots, 9)
\]
We make the SO(5,1)$\times$SO(4) decomposition of 
the 10D gamma matrices as follows.
\EQA
&&\Gamma^\mu=\gamma^\mu \otimes \hat{\gamma}\qquad 
(\mu=0,5,\ldots,9) \nonumber\\
&&\Gamma^m =\identity \otimes  \hat{\gamma}^m\qquad
(m=1,\ldots, 4)
\label{10Dgamma}
\EQN
where we have defined $\hat{\gamma}=
\hat{\gamma}^1\hat{\gamma}^2\hat{\gamma}^3\hat{\gamma}^4$.
The SO(5,1) gamma matrices $\gamma^\mu$ 
have 8$\times$8 components and the SO(4) gamma matrices
$\hat{\gamma}^m$ have 4$\times$4 components.
We further assume the explicit representation for
the SO(4) gamma matrices
\begin{equation}
\hat{\gamma}^m=\left(\begin{array}{cc} 0 & \sigma^m \\ 
-\bar{\sigma}^m &0 \end{array}\right)
\label{4Dgamma}
\end{equation}
where $\sigma_i$ ($i=1,2,3$) are Pauli matrices,
and $\sigma_4 =i\identity$. We also define 
$\bar{\sigma}_i=
-\sigma_i$ and $\bar{\sigma}_4=\sigma_4$.

The chirality in 10D is the product of 6D and 4D
chiralities, and in the above representation of 
gamma matrices,
\[
\bar{\Gamma}^{(10)}=\overline{\gamma}\otimes
\left(\begin{array}{cc} \identity_{2\times 2} &0 \\0 & 
-\identity_{2\times 2}
\end{array}\right)
\]
where $\overline{\gamma}=\gamma^0\gamma^5\cdots\gamma^9$ is
the 6D chirality matrix. Consequently, 10D Weyl spinor
$\Theta$
is decomposed into two pairs of 6D Weyl spinors
\[
\Theta =\left(\begin{array}{c}\theta_{+,1}\\ 
\theta_{+,2}\\
\theta_{-,1}\\ \theta_{-,2}
\end{array}\right).
\]

The matrix $B^{(10)}$ also allows the SO(5,1)$\times$SO(4)
decomposition 
\[
B^{(10)}=B^{(6)}\otimes B^{(4)}
\]
where $B^{(6)}$ and  $B^{(4)}$ satisfy
\[
B^{(6)}\gamma^\mu B^{(6)}{}^{-1}=-\gamma^\mu{}^*,\qquad
B^{(4)}\hat{\gamma}^m B^{(4)}{}^{-1}=-\hat{\gamma}^m{}^*.
\]
We note here the relation for $B^{(6)}$
\[
B^{(6)}{}^T=-B^{(6)},\qquad B^{(6)}{}^*=-B^{(6)}{}^{-1}
\]  
In the text, we omit the superscript (6) on $B^{(6)}$.

In the representation (\ref{4Dgamma}), $B^{(4)}$ is given
by
\[
B^{(4)}=\left(\begin{array}{cc} \epsilon &0 \\ 0&\epsilon
\end{array}\right)\quad \mbox{where} \quad
\epsilon=\left(\begin{array}{cc} 0 & 1 \\ -1 &0\end{array}\right),
\]
thus, 10D Majorana condition (\ref{MWcond}) in 10D becomes
the SU(2) Majorana condition in 6D.
\[
\Theta^\dagger =\left(\begin{array}{c}\theta_{+,1}^\dagger\\ 
\theta_{+,2}^\dagger \\
\theta_{-,1}^\dagger\\ \theta_{-,2}^\dagger\end{array}\right)
=\left(\begin{array}{c}B^{(6)}\theta_{+,2}\\ -B^{(6)}\theta_{+,2}\\
B^{(6)}\theta_{-,2}\\ -B^{(6)}\theta_{-,1}\end{array} \right)
\]

Since we prefer using unconstrained fermions
to perform the loop calculations, we explicitly 
eliminate half of the components of $\Theta$
($\theta_{+,2}$ and $\theta_{-,2}$)
using the SU(2) Majorana conditions. That is,
we take
\EQ
\Theta=\left(\begin{array}{c}\theta_{+}\\ 
-B^{(6)}{}^*\theta_{+}^\dagger \\
\theta_{-}\\ -B^{(6)}{}^* \theta_{-}^\dagger\end{array}\right).
\label{Theta10D6D}
\EN
The fermionic part of the action in (\ref{S0}) 
is obtained from the 10D covariant form
\[
{1\over g_s\ell_s}\int dt \Tr \Big({i\over 2}\overline{\Theta}
\Gamma^0 D_0\Theta 
+{1\over 2\lambda}\overline{\Theta}\Gamma^i [\Theta, X_i] \Big)
\]
by taking the above representation of gamma matrices
and substituting (\ref{Theta10D6D}).

\section{Derivation of the SUSY invariance of the
one-loop effective action}
The one-loop effective action of the BD 
Matrix theory is given by integrating out the
0-4 sector fields from the classical action 
$S_0[\phi]+S_5[\phi,\varphi]$.
\EQ
e^{-(S_0[\phi]+S^{(1)}[\phi])} =\int {\cal D} \varphi\,
e^{-(S_0[\phi]+S_5[\phi,\varphi])}
\label{defEA}
\EN
Note that we denote the 0-0 sector fields by $\phi$ and
0-4 sector fields by $\varphi$, and that the integration
$\int\! d\tau$ is implicit throughout this appendix.

As a consequence of the  SUSY invariance of the classical action, 
the following Ward identity holds:
\EQ
0=\int\! {\cal D}\varphi\,
 \left(\delta_\eta \phi {\delta S_0[\phi] \over \delta \phi}
+\delta_\eta \phi {\delta S_5 [\phi,\varphi]\over \delta \phi}
+\delta_\eta \varphi {\delta S_5 [\phi,\varphi]\over \delta \varphi}
\right)e^{-(S_0[\phi]+S_5[\phi,\varphi])}
\label{WI}
\EN
where $\delta_\eta \phi$ and $\delta_\eta \varphi$ are
the classical SUSY transformations given in 
(\ref{BDSUSY00})-(\ref{BDSUSY04}).
We shall denote the part of $\delta_\eta \phi$ which 
contain only the 0-0 sector fields by $\delta^{(0)}_\eta \phi$,
and the part containing 0-4 sector fields by
$\delta'_\eta \phi$.

We first note that the last term of (\ref{WI})
vanishes, for it is
the infinitesimal form of the 
invariance of the integral under the change of variables
\[
\int\! {\cal D}\varphi\, e^{-S_5[\phi,\varphi]}=
\int\! {\cal D} (\varphi+\delta\varphi)\, 
e^{-S_5[\phi,\varphi+\delta\varphi]}=
\int\! {\cal D}\varphi \,
e^{-S_5[\phi,\varphi+\delta\varphi]}
\]
where we have taken into account the invariance of 
the measure.

The first term of (\ref{WI})
gives the variation of $S_0$ with respect to the
expectation value of the classical SUSY
transformation $\delta \phi$ (times $e^{-(S_0+S^{(1)})}$)
\[
\int\! {\cal D}\varphi\,
\delta_\eta \phi {\delta S_0\over \delta \phi}
e^{-(S_0+S_5)}
= (\delta^{(0)}_\eta \phi+ \delta^{(1)}_\eta \phi)
{\delta S_0[\phi]\over \delta \phi} 
e^{-(S_0[\phi]+S^{(1)}[\phi])}
\]
where 
\EQ
\delta^{(1)}_\eta \phi\equiv
{\int\! {\cal D}\varphi\, \delta'_\eta \phi\,
e^{-S_5} \over \int\! {\cal D}\varphi\,
e^{-S_5}}.
\EN

Finally, if we evaluate the second term of (\ref{WI})
{\it to the  one-loop order}, it reduces to 
\EQ
\int\! {\cal D}\varphi\,
\delta_\eta \phi {\delta S_5[\phi,\varphi] \over \delta \phi}
e^{-(S_0[\phi]+S_5[\phi,\varphi])}
= \delta^{(0)}_\eta \phi \int\! {\cal D}\varphi\,
 {\delta S_5[\phi,\varphi] \over \delta \phi}
e^{-(S_0[\phi]+S_5[\phi,\varphi])}.
\label{RHSsecond}
\EN
For the present model,
\EQ
\int\! {\cal D}\varphi\,
\delta'_\eta \phi
{\delta S_5[\phi,\varphi] \over \delta \phi}
e^{-(S_0[\phi]+S_5[\phi,\varphi])}
\label{RHSsecond2}
\EN
which would be present in the RHS of (\ref{RHSsecond})
has at least two loops, for 
\[
\delta'_\eta \phi
{\delta S_5[\phi,\varphi] \over \delta \phi}
\]
is a four-point vertex as we see from the explicit
forms of $\delta'\phi$  (\ref{BDSUSYthmp}) and
$S_5$ (\ref{S5}).
We further note that the RHS of (\ref{RHSsecond}) 
gives the variation of the one-loop effective action 
(times $e^{-(S_0+S^{(1)})}$) as we see from
\[
{\delta S^{(1)} \over \delta \phi} =
{\int {\cal D}\varphi
 {\delta S_5\over \delta \phi}
e^{-S_5} \over \int {\cal D}\varphi
e^{-S_5}}
\]
which results from the definition (\ref{defEA}) of the
effective action.

Taking these facts into account, to the one-loop order,
(\ref{WI}) is rewritten as 
\EQ
0=\delta^{(0)}_\eta \phi {\delta S_0[\phi]
\over \delta \phi} +  \delta^{(1)}_\eta \phi {\delta S_0[\phi]
\over \delta \phi}
+\delta^{(0)}_\eta \phi {\delta S^{(1)}[\phi]\over \delta \phi}
\EN
which shows that the effective action 
$S_0+S^{(1)}$ is invariant
under effective SUSY transformation 
$\delta^{(0)}_\eta \phi+\delta^{(1)}_\eta \phi$. 

We note that the present discussion is not directly
applicable for showing the SUSY of 
the effective action  at 2-loop order and beyond. 
At these orders, 
(\ref{RHSsecond}) should no longer be valid,
due to the contribution from (\ref{RHSsecond2}).

In addition, we remark here
that our simple discussion which involve
only the Ward identity for the SUSY is due to the fact
that the gauge fields (which belong to the 0-0 sector)
are not integrated. When they are to be integrated, as
in the case of the BFSS model, analysis of the Ward identity
for the SUSY plus the one for the gauge symmetry is required
for studying the invariance of the effective action
under effective SUSY.
It is essentially due to the fact that the gauge fixing term
and the ghost action breaks SUSY. 
The formalism is developed in ref.\cite{KM1} and explicit form
of the effective SUSY transformation laws for the
one-loop effective action of BFSS model (for abelian v.e.v.) 
is given in ref.\cite{KM2}.

\end{document}